\documentclass{PoS}
\def\beq{\begin{equation}}
\def\eeq{\end{equation}}
\def\beqa{\begin{eqnarray}}
\def\eeqa{\end{eqnarray}}
\def\ba{\begin{array}}
\def\ea{\end{array}}

\def\a{\alpha}

\title{Gauge Unification from Split Supersymmetric String Models}

\ShortTitle{}

\author{{Christos Kokorelis}\\
         Physics division, National Technical University of Athens, Zografou campus, 15780, Athens, Greece $\&$ Hellenic Military Academy, Vari-Athens, 16673, Greece\\
        E-mail: \email{kokorelis@central.ntua.gr}}

\abstract{We discuss the unification of gauge coupling constants in non-supersymmetric open string vacua that possess the properties of Split Supersymmetry, namely the Standard Model with Higgsinos at low energies and where the
Standard model spectrum is always accompanied by right handed neutrinos. 
These vacua achieve partial unification of two out of three (namely SU(3)$_c$, SU(2), U(1)) running gauge couplings, possess massive gauginos and light Higgsinos at low energies and also satisfy $sin^2\theta_w (M_s) = 3/8$.
These vacua are based on four dimensional orbifold $Z_3 \times Z_3$ compactifications of  string IIA orientifolds with D6-branes intersecting at angles, where the (four dimensional) chiral fermions of the Standard Model appear as opens strings streching between the intersections of seven dimensional objects the so called D6-branes.} 

\FullConference{Proceedings of the Corfu Summer Institute 2015 "School and Workshops on Elementary Particle Physics and Gravity"\\
		1-27 September 2015\\
		Corfu, Greece}

\begin{document}

\section{Introduction}

 Recently,  ATLAS and CMS experiments
shed light on Higgs properties by combining the 2011 $\&$
2012 results from LHC run 1. Their
results in May 2015 
clearly show an improvement in precision (see also talk by G.Tonelli
in the school) of the experiments determing the Higgs mass (see figure 1 - taken from \cite{tonelli}) .
\begin{figure} 
\begin{center}
\includegraphics[width=.7\textwidth]{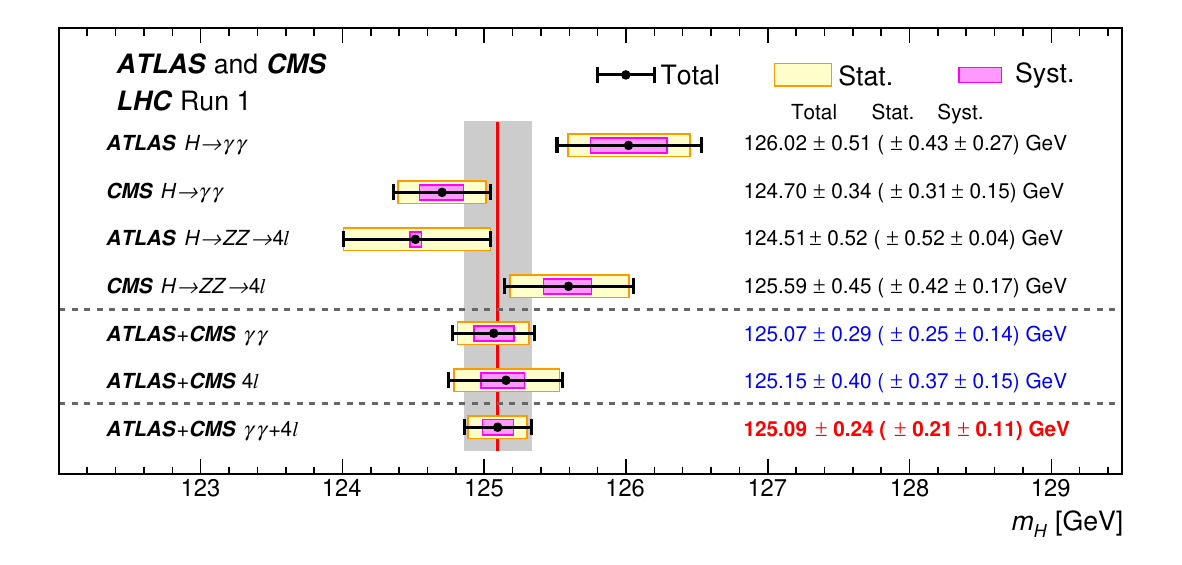}
\end{center}
\caption{Summary of Higgs boson mass measurements from the individual analyses of ATLAS
and CMS and from the combined analysis . The systematic (narrower,
magenta-shaded bands), statistical (wider, yellow-shaded bands), and total (black error bars)
uncertainties are indicated. The (red) vertical line and corresponding (gray) shaded column
indicate the central value and the total uncertainty of the combined measurement, respectively.
{\em The value of the Higgs mass is well below the value supported from the Standard model (SM),  thus suggesting the 
existence of another framework where the  SM is embedded, that can be a string theory with a high (as the one we are discussing in this talk) or a 
low string scale (e.g. from extra dimensions).} }
\end{figure}
The value of the Higgs mass is well below the value supported from the Standard model (SM),  thus suggesting the 
existence of another framework where the  SM is embedded, that can be a string theory with a high or a 
low string scale (e.g. from extra dimensions). The SM is an incomplete theory for the following reasons : 
1) does not incorporate gravity,  2) does not describe dark matter and dark energy. In fact,  
cosmological observations tell us that the standard model explains about 5$\%$ of the matter present in the universe.  
  About $\approx 27 \%$ should be dark matter, which can behave just like other matter, but which only interacts weakly (if at all) with the Standard Model fields
and also 3)  SM does not predict a  mass for the neutrinos.  At the SM m$_{\nu} =$ 0, but 
measurements however indicated that neutrinos spontaneously change flavour, which implies that neutrinos have a mass. This necessitates an extension of the standard model, which not only needs to explain how neutrinos get massive, but also why their mass is so small.
The solution to these problems is that the SM may be extended.\newline
\subsection{\em{High Scale N=1 Supersymmetry}} 
\begin{figure} 
\begin{center}
\includegraphics[width=.7\textwidth]{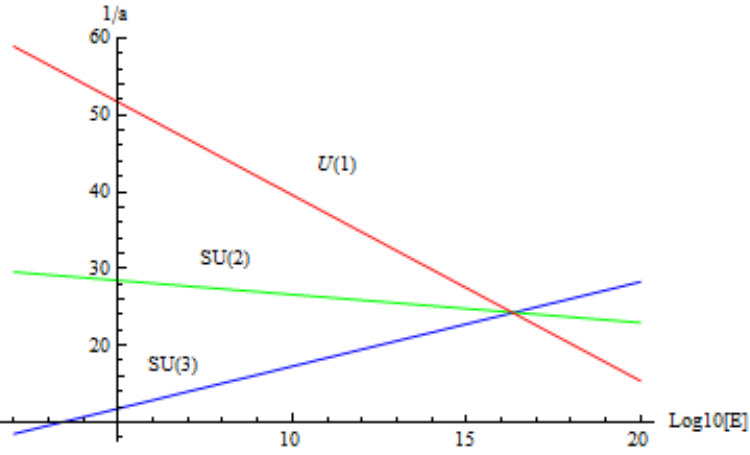}
\end{center}
\caption{Unification of running gauge coupling constants in the MSSM at tree level}
\label{figu}
\end{figure}

The first possibility is to use $N=1$ Supersymmetry (SUSY) in gauge theories
to promote the SM to a Supersymmetric Standard Model (SSM), e.g. to the so called Minimal Supersymmetric Standard Model (MSSM) \cite{dimogeo},
 such that for every Standard Model particle    
   there exists its supersymmetric (susy) partner (s-particle partner) with the same charge but with different spin  (a boson acquires an s-fermion susy partner and a fermion an s-boson susy partner). 
N=1 SUSY solves the gauge hierarchy problem and the running gauge couplings for the $SU(3)_c$, $SU(2)$ $U(1)_Y$ gauge group,  corresponding to  the strong, weak and hypercharge gauge couplings, unify \cite{dimo} at the unification scale \cite{unify} of 
\begin{equation}
 M_{GUT} \approx 2 \cdot 10^{16} \  GeV ,
 \label{gau1} 
\end{equation}
as seen in figure (\ref{figu}).
In a gauge theory context, there is no fundamental reason from first principles why the unification scale should be so high, if not by accident.
 However in N=1 supersymmetric vacua arising from the (perturbative) Heterotic String Theories
 the unification string scale $M_{string}$ is fixed to be, larger than its gauge theory
 counterpart (\cite{stringhe}) in eqn.(\ref{gau1}),
            \begin{equation}
M_{string}\stackrel{def}{=}\frac{2e^{(1-\gamma)/2} 3^{-3/4}}{\sqrt{2 \pi \alpha^{\prime}} } \approx 0.7 \ g_{string} \ 10^{18} \ GeV
\end{equation}                 
It appears, though, that $M_{string}$ is two (2) orders of magnitude larger than the unification of the SUSY gauge theory scale $M_{GUT}$.   
Several explanations were invoked to reconcile the apparent discrepancy, to name afew :  threshold corrections, extra states etc.(e.g. see \cite{lace}).
 
\section{D-branes in String Theory and the Gauge Couplings constant problem}

\begin{figure} 
\begin{center}
\includegraphics[width=.7\textwidth]{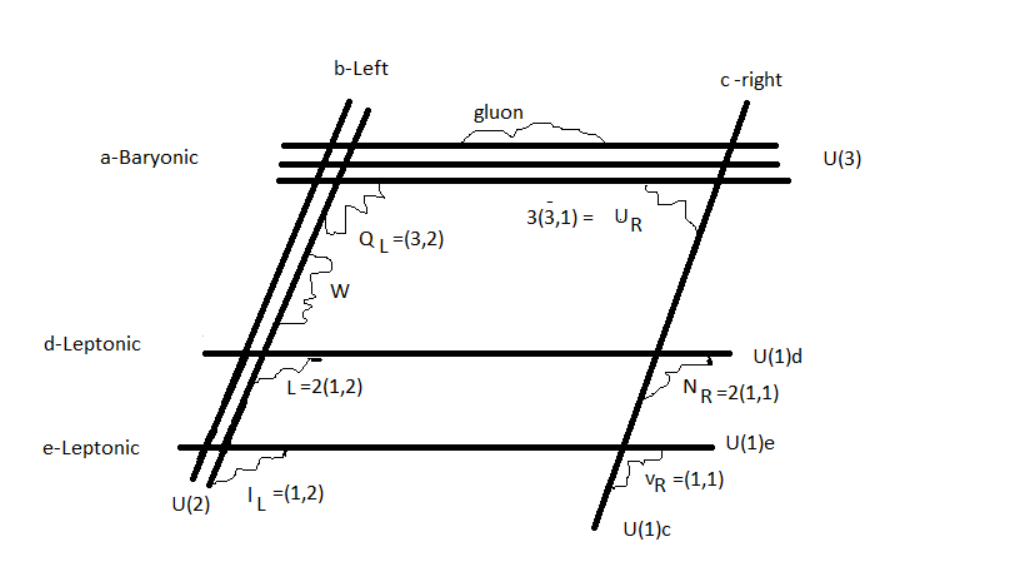}
\end{center}
\caption{The stringy Standard model includes the chiral fermions (CF) of the Standard Model and the right handed neutrinos. CF appear as open strings stretching between intersecting D6-branes in the 5-stack model of \cite{kokorelis1}. Baryon number is conserved as the corresponding gauge 
boson receives a string scale mass.
This model predicts an extra Z\rq{} gauge boson between 3.5-5.5 TeV \cite{bsg} and also 
accommodates the recent anomalies observed in the $ b \rightarrow s l^{+} l^{-}$ transitions observed by the LHCb
collaboration.}
\label{figura1}
\end{figure}

The above problems could be solved if we focus our attention to a string theory where the string scale is a free parameter
and could be lowered) so that it can be closer to the GUT scale of  eqn.(\ref{gau1}).
This goal can be achieved in theories which are compactifications of type I theories, namely compactifications of 10 dimensional  
IIA orientifolds with D6-branes intersecting at angles \cite{lust, kokore1}. On such string vacua \cite{imr}, \cite{kokorelis1}, \cite{kokos6}, \cite{report}, supersymmetry is already broken at the string scale, by construction, and   
at low energies the (chiral spectrum of the) SM survives to low energies. In these constructions, better known as the models of the Madrid group \cite{imr}, \cite{kokorelis1}, the SM appears in a global configuration and subsequently gets localized in a string model construction, by satisfying the ultraviolet tadpole constraints (the cancellation of cubic gauge anomalies) of the underlying string theory. 
In these constructions, chiral fermions of the SM appear as open strings stretching between intersecting D6-branes as in fig. (\ref{figura1}). In (\ref{figura1}) the five stack SM model of \cite{kokorelis1} is depicted. The gauge group of the model is the SM one, $SU(3)_c \times SU(2)_w \times U(1)_Y$. All extra U(1)\rq{}s, beyond the hypercharge  that are originally present at the string scale gets a mass from their non-zero couplings to Green-Schwarz couplings or some other mechanism.   
The chiral spectrum of the
model may be obtained by solving simultaneously the intersection constraints (that determine the multiplicity of the chiral fermions) coming
from the existence of the different sectors and the RR tadpole cancellation conditions. The SM chiral spectrum of the SM get its mass from a set of two-Higgses and the right handed neutrino gets a  Dirac mass.
In this model, by construction the string/unification can be found at a high scale while Baryon number survives at low energies as a global symmetry by construction, avoiding the baryon number decay problems of heterotic orbifold contructions/F-theory constructions. 
It has been shown recently \cite{bsg}, that this model explains the latest CERN
particle physics decay experiment $b \rightarrow s\ l^{+} \ l^{-}$  and predicts a new boson Z' with non-negligible
couplings to the first two quark generations and a mass in the range [3.5, 5.5] TeV with the possibility
to discover such a state directly during the next LHC runs via Drell-Yan production in the di-electron
or di-muon decay channels.

\subsection{Split Susy in String Theory - Explicit realizations }

Split SUSY in gauge theories (from now on named global Split Susy) \cite{ArkaniHamed:2004fb} has a number of features :  \newline
1) does not solve the Gauge Hierarchy problem but keeps the unification and dark matter candidate, \newline 2) it achieves  partial unification of two out of three SM gauge couplings at $\approx 10^{16}$ GeV,  \newline 3) All superpartners are considered massive, \newline 4) Higgsinos, gauginos remain at low energy, \newline 5) there is a light and a heavy Higgs and the SM (assumed to exists as a result of fine tuning). \newline
Split supersymmetry in string theory on the other hand  was introduced during November 2004 in \cite{anto}, \cite{kokore}. In \cite{anto} where global (quiver) models were constructed, it was argued that {\em String Theory can achieve Split Susy} if a number of conditions are satisfied, namely :\newline
i) $ sin^2 \theta_w = 3/8 $ at the string unification scale $M_s$ \newline
ii) there is partial unification of two (2) out of three (3) gauge couplings of the SM, \newline
 iii) there are light gauginos and iv) there are light Higgsinos, surviving to low energies \newline
  The 5th condition of global Split Susy did not appear in the present models of string theory.
In our work \cite{kokore} we proposed that for intersecting D6-brane Split Susy models, gauginos could get massive from loop corrections while specific string models with $sin^2 \theta = 3/8 $ at $M_s$ were proposed, arising from new four dimensional constructions \cite{kokore1} based on chiral four dimensional T6/($Z3 \times Z3$) orientifold compactifications of IIA theory that satisfy the criteria i) - iv).

\section{Non-supersymmetric Standard Model Gauge unification in Split Susy string models}

Our orientifold constructions \cite{kokore1} are based on IIA theory compactified on the 
${\bf T^6/(Z_3 \times Z_3)}$ 
orbifold, where the latter symmetry is generated by the twist
generators (where $\a = e^{\frac{2\pi i}{3} }$)
$\theta : (z_1, z_2, z_3) \rightarrow (\a z_1, \a^{-1} z_2, z_3) $,  
$\omega :(z_1, z_2, z_3) \rightarrow ( z_1, \a z_2, \a^{-1}z_3)$,
where $\theta $, $\omega$ get associated to the twists 
$\upsilon =  \frac{1}{3}(1, -1, 0)$, 
$ u = \frac{1}{3}(0, 1, -1)$.  Here, $z^i = x^{10-2i}+ i x^{11-2i}$, 
$i=1,2,3$ are the complex coordinates on the  
$T^6$, which we consider as being factorizable for simplicity, e.g.
$T^6 = T^2 \otimes T^2 \otimes T^2$. 
In addition, to the orbifold action 
the IIA theory is modded out by the orientifold 
action $\Omega R$ that 
combines the worldsheet parity  $\Omega $ and the antiholomorphic operation 
$R : z^i \rightarrow {\bar z}^i$.
and the 
$\Omega R $ action is along the horizontal directions across the six-torus. 
The model contains nine kinds of orientifold planes, that correspond to 
the orbit $\cal O$ consisting of the actions of $\Omega R$,  $\Omega R \theta$, $\Omega R \omega$,  
$\Omega R \theta^2$, $\Omega R \omega^2$, 
 $\Omega R \theta  \omega$,  $\Omega R \theta^2  \omega$, 
$\Omega R \theta  \omega^2$,  $\Omega R \theta^2  \omega^2$.
The closed
string spectrum contains gravitational multiplets and orbifold 
moduli and is not of any interest to us in the present review. In order to cancel the RR crosscap tadpoles introduced by the 
introduction of the orientifold planes we introduce N D6$_a$-branes of open strings wrapped
along three-cycles that are taken to be products of one-cycles along the 
three two-tori of the factorizable $T^6$. A D6-brane $a$ - 
associated with the equivalence class of wrappings $(n^I, m^I)$, $I=1,2,3$, - is mapped 
under 
the orbifold and orientifold action to its images
\beqa
a \leftrightarrow \left( \ba{c} n_a^1, m_a^1\\
 n_a^2 , \ m_a^2\\ n_a^3, \ m_a^3\ea \right), \ \theta a \rightarrow \left( \ba{c} 
-m_a^1 , \ (n-m)_a^1\\(m-n)_a^2 ,\ -n_a^2\\
n_a^3 ,\ m_a^3\ea \right), \ \ 
\Omega R a \rightarrow \left( \ba{c} (n-m)_a^1 ,\ -m_a^1\\
(n-m)_a^2 , \ -m_a^2\\ (n-m)_a^3 ,\ -m_a^3\ea \right) \ .
\eeqa
The $Z_3 \times Z_3$ orientifold models are subject to the cancellation of untwisted RR tadpole 
conditions \cite{kokore1} given by
 \beq
\sum_a N_a  Z_a = 4, 
\label{tad}
\eeq    
where
\beq
Z_a = 2 m^1_a m^2_a m^3_a + 2 n^1_a n^2_a n^3_a - n_a^1 n_a^2 m_a^3 - 
n_a^1 m_a^2 n_a^3 - 
m_a^1 n_a^2 n_a^3 -
m_a^1 m_a^2 n_a^3 -m_a^1 n_a^2 m_a^3 - n_a^1 m_a^2 m_a^3 
\eeq
The gauge group U($N_a$) produced by $N_a$ coincident D6$_a$-branes comes 
from the $a({\tilde a})$ sector, the sector made from 
open strings stretched between the $a$-brane and its images under the orbifold 
action. Also, we get three adjoint N=1 chiral multiplets. 
In the $a({\cal O}b)$ sector - strings stretched between the brane $a$ and the 
orbit images of brane $b$ - will localize $I_{ab}$ fermions in the 
bifundamental $(N_a, {\bar N}_b)$ where 
\beq
I_{ab}= 3(Z_a Y_b - Z_b Y_a),
\eeq 
and $(Z, Y) $ are the effective wrapping numbers with $Y_a$ given by
\beq
Y_a = m^1_a m^2_a m^3_a + n^1_a n^2_a n^3_a - n^1_a n^2_a m^3_a - 
n^1_a m^2_a n^3_a - m^1_a n^2_a n^3_a 
\eeq
The sign of $I_{ab}$ denotes the chirality of the associated fermion, where
we choose positive intersection numbers for left handed fermions. 
In the sector ${ab^{\prime}}$ - strings stretching between the brane $a$ and
the orbit images of brane $b$, there are $I_{ab^{\prime}}$ chiral fermions 
in the bifundamental $(N_a, N_b)$, with
\beq
I_{ab^{\prime}}= 3(Z_a  Z_b - Z_a Y_b - Z_b Y_a),
\eeq 
Chiral fermions
in symmetric (S) and antisymmetric (A) representations of $U(N_a)$ from open
strings stretching between the brane $a$ and its orbit images $({\cal O}a)$ appear as, 
\beqa
(A_a) = 3(Z_a - 2 Y_a) ,\\
(A_a + S_a) = \frac{3}{2}(Z_a - 2 Y_a)(Z_a - 1)
\eeqa 
Also, from open strings stretched between the brane $a$ and its orbifold 
images we get non-chiral massless fermions in the adjoint representation,
\beq
(Adj)_L : \prod_{i=1}^3 (L^I_{[a]})^2 \ ,
\label{adj1}
\eeq
where 
\beq
L^I_{[a]} = \sqrt{(m_a^I)^2 + (n_a^I)^2 -(m_a^I) (n_a^I) }
\label{nonma}
\eeq 
Adjoint massless matter, including fermions and gauginos that are massless at 
tree level 
are expected to receive string scale masses from loops once supersymmetry is broken \cite{imr}.
The evolution of the one loop renormalization group equations (ERGE) for the $SU(3)_c$, $SU(2)_w$, $U(1)_Y$ gauge couplings in the absence of one-loop string threshold corrections  
(see \cite{lsti} and also \cite{kokore1}) are given by 

\beqa
\frac{1}{a_s (M_z)} = \frac{1}{a_s (M_s)} - \frac{b_3}{2 \pi} \ln \left( \frac{M_Z}{M_s}\right), \nonumber\\
\frac{sin^2 \theta_w (M_Z)}{a_{em} (M_Z)} = \frac{1}{a_w (M_s)} -\frac{b_2}{2\pi} \ln \left(\frac{M_Z}{M_s}\right),\nonumber\\
\frac{cos^2\theta_w (M_Z)}{a_{em} (M_Z)} = \frac{1}{a_Y (M_s)} -\frac{b_1}{2\pi} \ln \left( \frac{M_Z}{M_s}\right),
\label{RG}
\eeqa
\beq
\frac{2}{3} \frac{1}{ \alpha_s (M_Z)}  + \frac{2 sin^2 \theta_w (M_Z) - 1}{\alpha_{em} (M_Z)} =\frac{B}{2 \pi} \ \ln \frac{M_Z}{ M_s} \ ,
\label{strings}
\eeq
and $B=-(\frac{2}{3} b_3 + b_2 - b_Y)$; $b_3$, $b_2$, $b_1$ are the $\beta$-function coefficients for strong, 
weak and hypercharge gauge couplings respectively. The string scale gets calculated from eqn. (\ref{strings}).
For a theory which accommodates the SM and a number of extra particles below a scale $M_s$
\beq
 ( b_1, \ b_2 , \ b_3 ) \ = \ ( \frac{20}{9}n_G + \frac{1}{6}n_H + N_1,\   \frac{4}{3}n_G + \frac{1}{6}n_H + N_2 -\frac{22}{3},   \ 
\frac{4}{3}n_G -11 + N_3),  
\label{betas}
\eeq
where $N_1, N_2, N_3$ the contribution of the beyond the SM particles; the rest of the terms in 
(\ref{betas}) are the Standard model contributions; $n_G$ the number of generations; $n_H$ 
the number of Higgses. 
{\em It appears \cite{kokore} that in a non-supersymmetric string model, the value of the string scale which depends on the variable B is independent of the number of Higgses as their dependence cancels out. 
In table (\ref{flotab1}), we can clearly see that the string unification scale, if we have either the SM with right handed neutrinos at low energy or the SM accompanied by 3 pairs of hisggsinos, the string scale value is fixed at $M_s = 5.1 \times 10^{13}$ Gev independent of the number of Higgses present as their dependence cancels out  in the combination $b_2 - b_1$ of } B.   
 
\begin{table}[htb] \footnotesize
\renewcommand{\arraystretch}{1}
\begin{center}
\begin{tabular}{|r|c|c|} 
\hline\hline
 ${\bf Spectrum \ at \ low \ energies\hspace{2cm}}$ & $\beta_{coefficents} \hspace{2cm}$ & $M_s$ \\
\hline\hline
$SM $   & $ (b_1, b_2, b_3)=(\frac{20}{3}+\frac{1}{6}n_H, -\frac{10}{3}+\frac{1}{6}n_H, -7), \ B=\frac{44}{3}$ & $ M_s \approx 5.1 \cdot 10^{13}$     \\\hline
$ \ SM + 3 \ pairs \  of \ H_u, H_d \ Higgsinos $   & $  (b_1, b_2, b_3)=(\frac{26}{3}+\frac{1}{6}n_H, -\frac{4}{3}+\frac{1}{6}n_H, -7), \ B=\frac{44}{3}$ & $M_s \approx 5.1 \cdot 10^{13}$  \\
\hline
\hline
\end{tabular}
\end{center}
\caption{\small
Value of the string scale $M_s$ in Split Sysy models. All models possess sin$^2 \theta_w = $3/8 at $M_s$.
\label{flotab1}}
\end{table}
The input values of the inverse gauge
couplings at low energy (i.e. at $m_Z$ = 91.1876 GeV ) for the SM have been fixed
by taking the recent PDG data \cite{pdg}
\begin{eqnarray}
\alpha^{-1}_1( m_Z) & = & 59.01 \pm 0.02\  ,\\
\alpha^{-1}_2 (m_Z) & = & 29.57 \pm 0.02 \ ,\\
\alpha^{-1}_3 (m_Z) & = &  \ \ 8.45 \pm 0.05 \ . \\ 
\end{eqnarray}

 $a_2 (M_Z) = a_{em} \ sin^2 \theta(M_Z)$,

\subsection{A Split Susy example}

The minimal choice of obtaining the SM gauge group and chiral spectrum 
originates from a three stack 
$U(3)_a \times U(2)_b \times U(1)_c$ gauge group with D6-branes intersecting at angles at the string scale \cite{kokore, kokore1}.  
The choice of wrapping numbers
\beq
(Z_a, Y_a) = \left( \ba{cc} 1,  & 0 \\\ea \right), \ (Z_b, Y_b) = 
\left( \ba{cc} 1, & 1 \\\ea \right), \ (Z_c, Y_c) = \left( \ba{cc}-1, & 1 \\\ea \right)
\label{wrap3}
\eeq 
satisfies the RR tadpoles (\ref{tad}) and corresponds to the spectrum seen in 
table (\ref{give1}).

The D-brane
model \cite{kokore} of table (\ref{give1}) has a non-supersymmetric spectrum. The Higgses come from vanishing intersections and can
be understood as part for the massive spectrum that organize itself in terms of massive
N=2 hypermultiplets. The Higgs fields become subsequently tachyonic in order to
participate in electroweak symmetry breaking (for a more detailed discussion of these issues see
\cite{imr, kokorelis1, kokos6}).
\begin{table}[htb] \footnotesize
\renewcommand{\arraystretch}{1.7}
\begin{center}
\begin{tabular}{|r|c|c|c|}
\hline\hline
 ${\bf Matter\hspace{2cm}}$ & Intersection & $(SU(3) \times SU(2))_{(Q_a, Q_b, Q_c)} \hspace{2cm}$ & $U(1)^{Y}$ \\
\hline\hline
$\hspace{2cm} \{ Q_L \} \hspace{2cm}$   & $ab*$ & $3({\bar{3}}, { 2})_{(-1,\ -1,\ 0)} \hspace{2cm}$ & $1/6$  \\
\hline
$\hspace{2cm}\{ u_L^c  \}\hspace{2cm}$   & $A_a$   & $3(3, 1)_{(-2,\ 0,\ 0)}\hspace{2cm}$ & $-2/3$  \\
\hline
$\hspace{2cm}\{ d_L^c \}\hspace{2cm}$ &  $ac$   & $3(3, 1)_{(1,\ 0, \  -1)}\hspace{2cm}$ & $1/3$ \\
\hline
$\hspace{2cm}\{ \ L \}\hspace{2cm}$ & $bc$  & $3(1, 2)_{(0,\  1,\  -1 )}\hspace{2cm}$ & $-1/2$ \\
\hline
$\hspace{2cm}\{ \ H_d \}\hspace{2cm}$ &bc$ $  & $3(1, 2)_{(0,\  1,\  -1 )}\hspace{2cm}$ & $-1/2$ \\
\hline
$\hspace{2cm}\{ H_u \}\hspace{2cm}$ & $bc*$   & $3(1, {\bar 2})_{(0,\  -1,\  -1 )}\hspace{2cm}$ & $1/2$ \\
\hline
$\hspace{2cm}\{ e_L^{+} \}\hspace{2cm}$ & $A_b $  & $3(1, 1)_{(0, \ -2,\ 0)}\hspace{2cm}$ & $1$ \\
\hline
$\hspace{2cm}\{ N_R \}\hspace{2cm}$ & $S_c $  & $9(1, 1)_{(0, \ 0, \ 2)}\hspace{2cm}$ & $0$ \\
\hline
\hline
$\hspace{2cm}\{ C_1 \}\hspace{2cm}$ & $ac $  & $3(3, 1)_{(1, \ 0, \ -1 )}\hspace{2cm}$ & $1/3$ \\
\hline 
$\hspace{2cm}\{ C_2 \}\hspace{2cm}$ & $ac* $  & $3({\bar 3}, 1)_{(-1, \ 0, \ -1 )}\hspace{2cm}$ & $-1/3$ \\
\hline
\hline
\end{tabular}
\end{center}
\caption{\small
A three generation 4D non-supersymmetric model with the chiral fermion content of 
N=1 Supersymmetric Standard Model on top of the table, in addition to $N_R$'s. There are three pairs
of $H_u$, $H_d$ Higgsinos. This model has the structure of models coming from gauge mediation scenarios and exhibits the spectrum of Split Supersymmetry with the SM and Higgsinos at low energy. The spartners are part of the
massive spectrum with mass of of order $M_s$.
and possess 
$sin^2 \theta = 3/8$ at $M_s$.  
\label{give1}}
\end{table}
All fermions become massive from the Yukawa couplings
\newpage
\beq
\lambda_d {(Q_L)} {(d^c_L)} \langle {{\tilde H}_d} \rangle \ + \ \lambda_{\nu} L N_R \langle {\tilde H}_u \rangle \  + \frac{\lambda_e}{M_s}L e_L^c \langle {\tilde H}_d \rangle   + (\lambda_H H_u H_d + \lambda_C C_1 C_2) \langle {\tilde N}_R \rangle,
\label{yuk}
\eeq
where the Higgs fields have the quantum numbers
\beq
{\tilde H}_u = (1, 2, 1)_{(0, -1, -1)}, \ \ {\tilde H}_d = (1, 2, 1)_{(0, 1, 1)} \ . 
\eeq
Lets us now examine gauge coupling unification using the models of table (\ref{give1}) as a representative example. They are non-supersymmetric but they respect 
\beq
\sin^2 \theta_W \stackrel{M_s}{=} \frac{3}{8} \  .
\label{sin}
\eeq
 as $a_2 = a_3$ at $M_s$. Thus at the 
unification scale where the gauge couplings $SU(2)_w$, $SU(3)_c$ meet, we have
the standard SU(5) GUT relation 
\beq
(5/3) a_Y = a_s = a_w = a_2 = a_3 
\eeq
This relation \lq{}solves\rq{} the following stringy relation among the gauge coupling constants
\beq
\frac{1}{a_Y}=\frac{2}{3}\frac{1}{a_s} +\frac{1}{a_w} \ .
\label{solv}
\eeq
For the Split Susy spectrum of model of table (\ref{give1}) we find
\beq
(b_1, b_2, b_3) = (\frac{29}{3}, -\frac{1}{3}, -7)
\eeq  
\begin{figure} 
\begin{center}
\includegraphics[width=.7\textwidth]{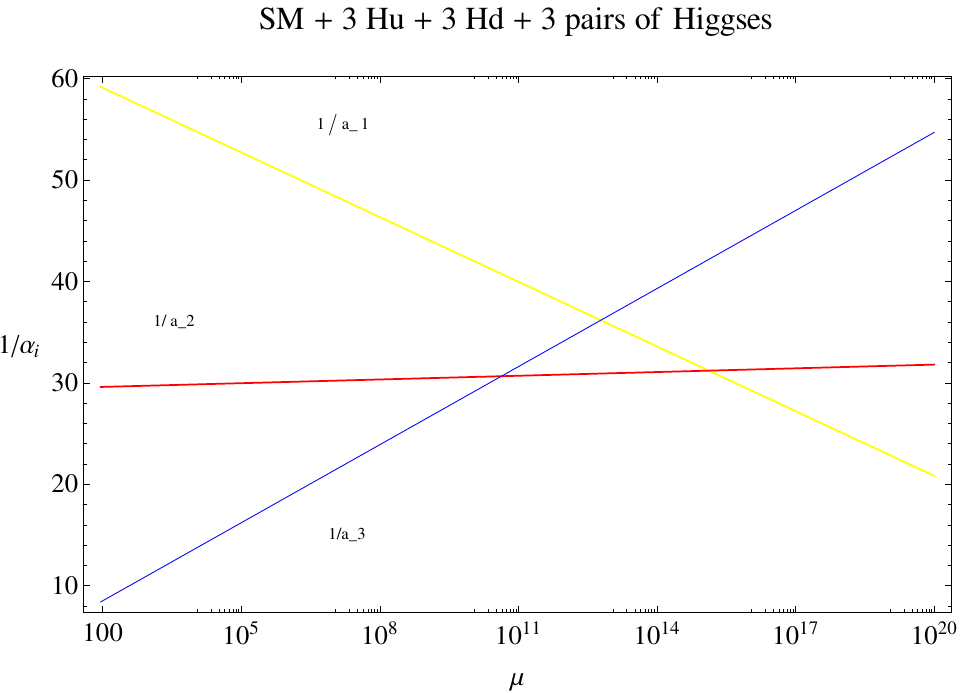}
\end{center}
\caption{Gauge coupling evolution with the energy for the stringy Split Susy Standard model.}
\end{figure}
We get a stringy version of unification of interactions, where
two out of three of gauge couplings unify.

\end{document}